\documentclass[conference]{IEEEtran}
\usepackage{tikz}
\usepackage{soul}
\usepackage{hyperref}
\usepackage{inconsolata}
\usepackage{graphicx}
\usepackage{xcolor}
\usepackage{comment}
\usepackage{lipsum}  
\usepackage[noadjust]{cite}
\usepackage{listings}
\usepackage{placeins}
\usepackage{tabularx}
\usepackage{amsmath}
\usepackage{amssymb}
\usepackage{multirow}
\usepackage{enumitem,pifont}
\usepackage{stfloats}
\usepackage{hhline}
\usepackage{rotating}
\usepackage{subcaption}

\usetikzlibrary{shapes.geometric, arrows.meta, shadows}
\tikzstyle{startstop} = [rectangle, rounded corners, minimum width=4cm, minimum height=1cm, text centered, draw=black, fill=red!30, drop shadow]
\tikzstyle{process} = [rectangle, minimum width=4cm, minimum height=1cm, text centered, draw=black, fill=blue!30, drop shadow]
\tikzstyle{arrow} = [thick,-Triangle, line width=1.5pt]

\IEEEoverridecommandlockouts
\IEEEpubid{\makebox[\columnwidth]{979-8-3503-5111-8/24/\$31.00~\copyright2024 IEEE \hfill} \hspace{\columnsep}\makebox[\columnwidth]{ }}

\begin{document} 

\title{Securing Satellite Link Segment:\\ A Secure-by-Component Design}

\author{%
\IEEEauthorblockN{Olfa Ben Yahia}
\IEEEauthorblockA{\textit{Polytechnique Montréal}\\
Montréal, QC, Canada \\
olfa.ben-yahia@polymtl.ca}
\and
\IEEEauthorblockN{William Ferguson}
\IEEEauthorblockA{\textit{303 Overwatch (A.S.B.L)}\\
Luxembourg City, Luxembourg \\
william.o.ferguson@ethicallyhacking.space}
\and
\IEEEauthorblockN{Sumit Chakravarty}
\IEEEauthorblockA{\textit{Kennesaw State University}\\
Kennesaw, GA, USA \\
schakra2@kennesaw.edu}
\and
\IEEEauthorblockN{Nesrine Benchoubane}
\IEEEauthorblockA{\textit{Polytechnique Montréal}\\
Montréal, QC, Canada \\
nesrine.benchoubane@polymtl.ca}
\and
\IEEEauthorblockN{Gunes Karabulut Kurt}
\IEEEauthorblockA{\textit{Polytechnique Montréal}\\
Montréal, QC, Canada \\
gunes.kurt@polymtl.ca}
\and
\IEEEauthorblockN{Gürkan Gür}
\IEEEauthorblockA{\textit{Zurich University of Applied Sciences}\\
Winterthur, ZH, Switzerland \\
gurkan.gur@zhaw.ch}
\and
\IEEEauthorblockN{Gregory Falco}
\IEEEauthorblockA{\textit{Cornell University}\\
Ithaca, NY, USA \\
gfalco@cornell.edu}
}

\maketitle

\IEEEpubidadjcol

\begin{abstract}

The rapid evolution of communication technologies, compounded by recent geopolitical events such as the Viasat cyberattack in February 2022, has highlighted the urgent need for fast and reliable satellite missions for military and civil security operations. Consequently, this paper examines two Earth observation (EO) missions: one utilizing a single low Earth orbit (LEO) satellite and another through a network of LEO satellites, employing a secure-by-component design strategy. This approach begins by defining the scope of technical security engineering, decomposing the system into components and data flows, and enumerating attack surfaces. Then it proceeds by identifying threats to low-level components, applying secure-by-design principles, redesigning components into secure blocks in alignment with the Space Attack Research \& Tactic Analysis (SPARTA) framework, and crafting \textsc{shall} statements to refactor the system design, with a particular focus on improving the security of the link segment.


\end{abstract}


\section{Introduction} \label{sec:intro}

 


Over the past few decades, satellites have become fundamental to our daily lives, often seamlessly integrating without notice. Each day, billions of people use satellite-enabled services for weather forecasts, telecommunications, and television. Specifically, Earth observation (EO) has expanded significantly since the 1960s, starting with the Tiros-1 satellite, which provided the first weather images from space \cite{achacheopen}. Today, EO spacecraft deliver essential imagery and remote-sensing data, supporting various applications with significant impacts across economic, societal, and environmental sectors. These applications include environmental monitoring, maritime surveillance, homeland security, land management, agriculture, meteorology, and disaster response management, demonstrating satellite technology's global significance and versatility \cite{crisp2020benefits}.


Traditionally, EO missions have relied on large, monolithic satellites equipped with heavy and complex payloads. These satellites are valued for their enhanced lifetime, coverage, and high-resolution imaging capabilities. However, recent advances from the New Space era in EO have proposed several small or medium-sized, low-cost satellite constellations. Their goal is to revolutionize operational and business models by offering cost-effective services that provide high-resolution and frequent revisits \cite{alandihallaj2022multiple}. The use of a single space vehicle for a LEO based solution is considered obsolete. Because this approach introduces a single point of failure and simply does not address the basic expectation of redundancy. In addition, the advent of CubeSat technology as an alternative greatly reduces the manufacturing costs in contrast to traditional monolithic EO solutions. In response, there has been a growing interest in using low Earth orbit (LEO) with satellite constellations. Operating satellites at this lower altitude allows smaller, less powerful payloads to achieve the same performance as their larger counterparts in higher orbits. Furthermore, there is a shift toward more flexible and resilient mission architectures, including both monolithic and distributed systems. In a monolithic system, a satellite performs observations and transmits data to Earth. Alternatively, distributed architectures involve multiple small satellites (fractionated spacecraft) that operate in coordination. This fractionated approach facilitates the easier replacement of individual modules if they fail, potentially extending the mission's overall lifetime~\cite{mathieu2006assessing}. This evolution in satellite technology and mission design represents a significant shift toward more efficient and scalable EO strategies~\cite{mccormick2009analyzing}.


With the development of LEO satellites for EO progressing rapidly, addressing their security vulnerabilities has become increasingly critical. EO missions are susceptible to various types of active and passive attacks, making each component within a mission a potential attack surface. In addition, the Viasat cyberattack in February 2022 \cite{viasat2022} highlighted the urgent need for reliable satellite missions for military and civilian security. Given the critical nature of these missions and the valuable data they collect, securing these components from the outset of the design process is essential. This proactive approach ensures that vulnerabilities are addressed before deployment, safeguarding the integrity and confidentiality of mission-critical information gathered from space.

In recent studies, numerous authors have explored the security aspects of EO and small satellites. The authors in \cite{murtaza2018simple} identified the limitations of the traditional ``store and later downlink" method and proposed a protocol that authenticates users for real-time image data access from satellites. Furthermore, the work in \cite{abdelaziz2019securing} examined the security vulnerabilities of space data link communication protocols for EO and suggested a software-based solution that uses software-friendly cryptographic algorithms. In addition, \cite{yue2023low} discussed the unique security and reliability challenges of LEO satellite communication systems due to their inherent characteristics. In another study~\cite{9697673}, the authors employed attack tree analysis to identify vulnerabilities in small satellites, specifically CubeSats. Meanwhile, the research in \cite{falco2019cybersecurity} delves into the cybersecurity weaknesses prevalent in the space sector and current mitigation techniques. This paper also proposes several security principles that address technical and policy issues to improve cybersecurity across all stakeholders. However, none of these papers have specifically addressed the concept of \textit{secure-by-design} from an EO perspective.


In contrast to existing research, this paper presents a novel approach by the authors, who are contributors to the IEEE SA P3349 Space System Cybersecurity Working Group.
The main key contributions in this work are as follows:
\begin{itemize}
    \item We present two use cases for EO missions, illustrating the design and decomposition of their components for enhanced security engineering purposes.
    \item We enumerate potential attacks in optical networks and identify potential attack surfaces from a link perspective.
    \item We demonstrate the application of the \textit{secure-by-component} methodology to the link segment in both scenarios.
\end{itemize}


\section{Earth Observation Use Cases} \label{sec:UseCases}






This section describes two EO use cases: standalone operations and integrating an inter-satellite link (ISL) with a computational offloading setup.

\subsection{EO through a Single LEO Satellite}

 \begin{figure}[!t]
   \centering
     \includegraphics[width=0.6\columnwidth]{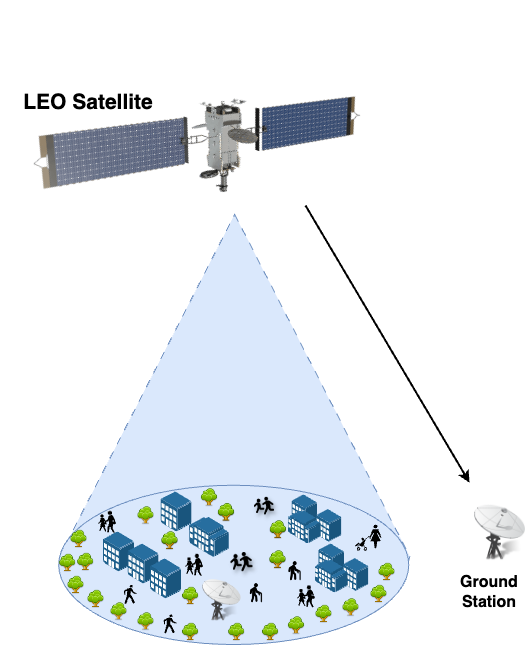}
   \caption{EO through a single LEO satellite.}
  \label{figure1}
 \end{figure}

EO satellites are designed to observe various events occurring on Earth from space. They are equipped with different sensors tailored to specific purposes, including monitoring natural phenomena, tracking disasters, and documenting changes to the Earth. In this subsection, we focus on a LEO satellite that observes Earth and transmits the collected data to a ground station, as depicted in Fig. \ref{figure1}. In this scenario, we utilize both uplink and downlink communications. The downlink primarily handles the transmission of collected data back to Earth, including images and sensor outputs, thus requiring higher data rates and bandwidth. Conversely, the uplink sends commands from Earth for parameter adjustments, necessitating lower data rates and frequencies. Both radio frequency (RF) and free-space optical (FSO) communications are viable options and will be discussed further in the following section.

\subsection{EO through a LEO Satellite Network}

 \begin{figure}[!t]
  \centering
       \includegraphics[width=0.87\columnwidth]{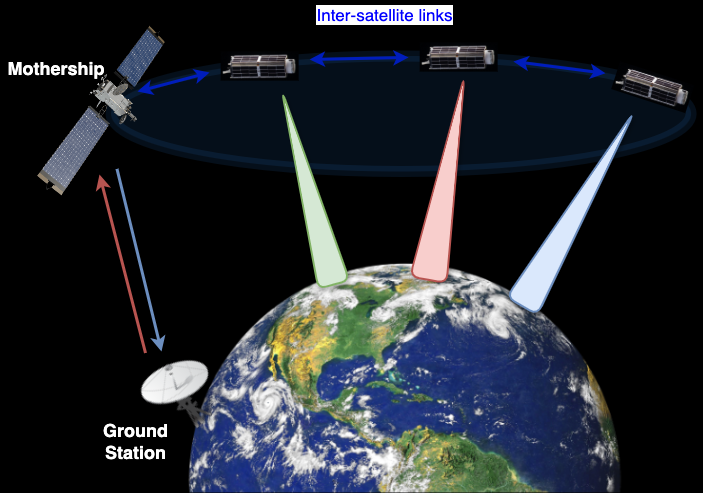}
    \caption{EO through a LEO satellite network.}
    \label{figure2} 
   \end{figure}

Multi-satellite constellations are gaining more attention because of their affordability and quick development time. Instead of using large, expensive satellites, we can deploy swarms of smaller satellites for various applications, including EO, atmospheric sampling, and synthetic apertures. This cost-effective approach can enhance the mission's capabilities. This subsection explores the New Space paradigm by considering a scenario with multiple small satellites. Specifically, we focus on a network of satellites interconnected via ISLs and connected to a mothership, which enhances EO capabilities. The satellites capture data, which is then transferred to the mothership. From there, it is transmitted to the ground station for further analysis, as depicted in Fig. \ref{figure2}. The mothership not only sends data back to Earth but also stores the collected data and controls the operations of the small satellites.

\begin{figure*}[htbp]
\centering
 \includegraphics[width=\textwidth]{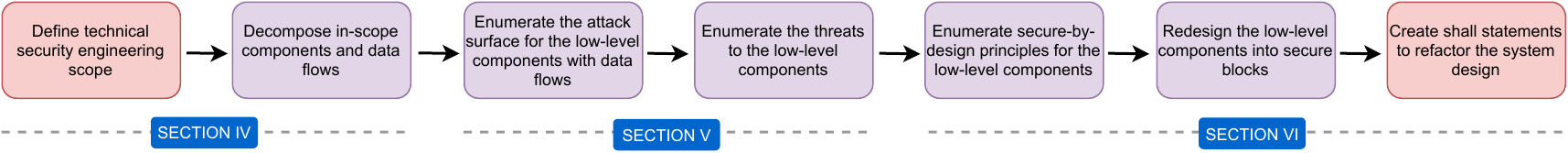}
\caption{\textit{Secure-by-component} design strategy.}
\label{figure:methodology}
\vspace{-1em}
\end{figure*}

\section{Security Challenges and Proposed Approach}\label{sec:SecureChallenges}

This section discusses the security challenges associated with the link side and outlines our proposed approach.


\subsection{Potential Vulnerabilities in Optical Links}


\begin{table}[t!]
\centering
\caption{Comparative table between RF and FSO}
\begin{tabular}{| m{1.4cm} | m{3cm} | m{3cm} |}
\hline
 {Feature} & {RF communication} & {FSO communication} \\
\hline\hline
{Medium} & Uses radio waves & Uses light waves\\
\hline
{Frequency range} & MHz to GHz & THz  \\
\hline
{Antenna size} & Much bigger antenna size due to much larger wavelength & Smaller antenna size due to smaller wavelength \\
\hline
{Weather sensitivity} & Less affected by weather conditions & Highly affected by weather conditions \\
\hline
{Power consumption} & Less received power for a given transmitted power & More received power for a given transmitted power\\
\hline
{Security} & Easy to intercept & More secure due to narrow beam and line-of-sight (LOS) requirement \\
\hline
\end{tabular}
\label{Table:compartive}
\vspace{-0.2cm}
\end{table}

Ensuring the security of communication channels is crucial when designing a network for EO, as these channels are susceptible to potential attacks. These channels are vital for transmitting the data gathered by the network. Two primary communication technologies are employed: RF and FSO. However, each technology has unique characteristics, security challenges, and vulnerabilities. Table \ref{Table:compartive} summarizes the main differences between RF and FSO communications.

Future satellite constellations are expected to increasingly rely on optical communication due to its higher data rates and inherent security features, as optical communication is generally considered more difficult to intercept. A study by \cite{kang2024survey} considers optical communication as a defensive scheme in satellite networks. However, optical networks are vulnerable to various attacks, including eavesdropping, high-power jamming, physical infrastructure attacks, denial of service, and tapping attacks that allow unauthorized access for eavesdropping and traffic analysis \cite{marudhai2022design, abdelsalam2023physical, 10004986, 9511459}. Most attacks on optical networks target the channel side, emphasizing the need for a \textit{secure-by-design} approach to the communication link, even when using FSO. For instance, \cite{9779543} demonstrated potential information leaks in satellite networking using optical communication. They proposed a model where a satellite communicates with a high-altitude platform station (HAPS) node through optical communication. In this model, an attacker positioned close to or on top of the satellite could eavesdrop on the transmitted signals. While inter-satellite communication is critical to the satellite system, its security remains underexplored.

Moreover, advances in FSO technology, while enhancing performance, introduce new security concerns. Attackers could inject signals to alter communications, degrading the signal-to-noise ratio (SNR) at the legitimate receiver and thus decreasing the overall secrecy capacity. Pointing errors and beam divergence caused by beam wander further exacerbate security risks by causing misalignment between the transmitter and receiver, lowering the main channel's adequate received power and SNR. Empirical data from NASA demonstrate that larger receiver apertures can reduce the bit error rate (BER) and aid in managing variable weather conditions. However, they also enable jammers to exploit this feature by transmitting disruptive optical pulses, making it difficult to differentiate legitimate signals from interference. This poses substantial risks to communication integrity, especially in high-security scenarios such as military applications \cite{marudhai2022design}.

This highlights the importance of designing secure communication systems from the outset to ensure the integrity and confidentiality of data transmission in optical communication networks. Special attention must be given to potential vulnerabilities, such as pointing errors and signal injection, which can degrade the main user channel and reduce the overall secrecy capacity. Additionally, beam divergence and alignment issues must be managed to maintain the reliability and availability of the communication link, particularly over long distances.

\subsection{Secure-by-Component Approach}\label{sec:SBD}

We implemented the \textit{secure-by-component} approach in \cite{Viswanathan2024} to address these challenges. This approach builds from secure building blocks to mission-level security and complements traditional top-down methods. We ensure a robust security framework by decomposing the system into secure components. Our methodology is mapped to specific sections of our paper, as shown in Fig.~\ref{figure:methodology}. Furthermore, we considered SPARTA \cite{aerospace2023sparta} to list attack techniques and \textit{secure-by-design} principles, though other frameworks can also be applied.

As a first step, we define our technical scope limited to the link segment, which encompasses the \textit{link capabilities} required for ground-to-space, space-to-ground, and, in the case of our EO scenarios, space-to-space communications. Applying this methodology to each segment separately is essential for a comprehensive security strategy to ensure thorough security coverage.

\section{Decomposition of In-Scope Components and Data Flows}\label{Components}

This section decomposes the two use cases into functional blocks to analyze the associated data flows.


\begin{figure*}[htpb!]
\begin{subfigure}[b]{0.45\textwidth}
    \centering
    \includegraphics[width=0.9\columnwidth]{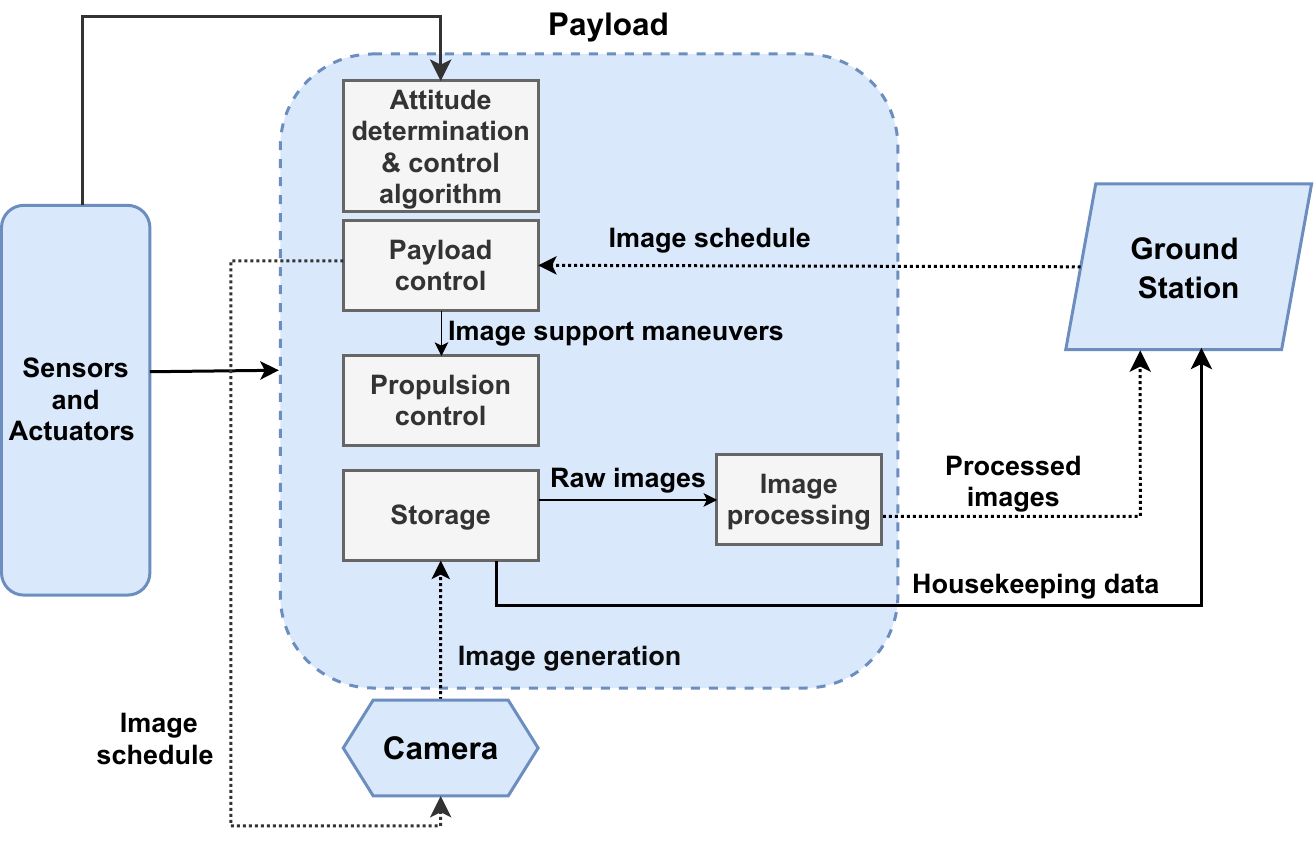}
    \caption{}
    \label{figure_single_leo}  
\end{subfigure}
\hfill
\begin{subfigure}[b]{0.45\textwidth}
\centering
 \includegraphics[width=0.9\columnwidth]{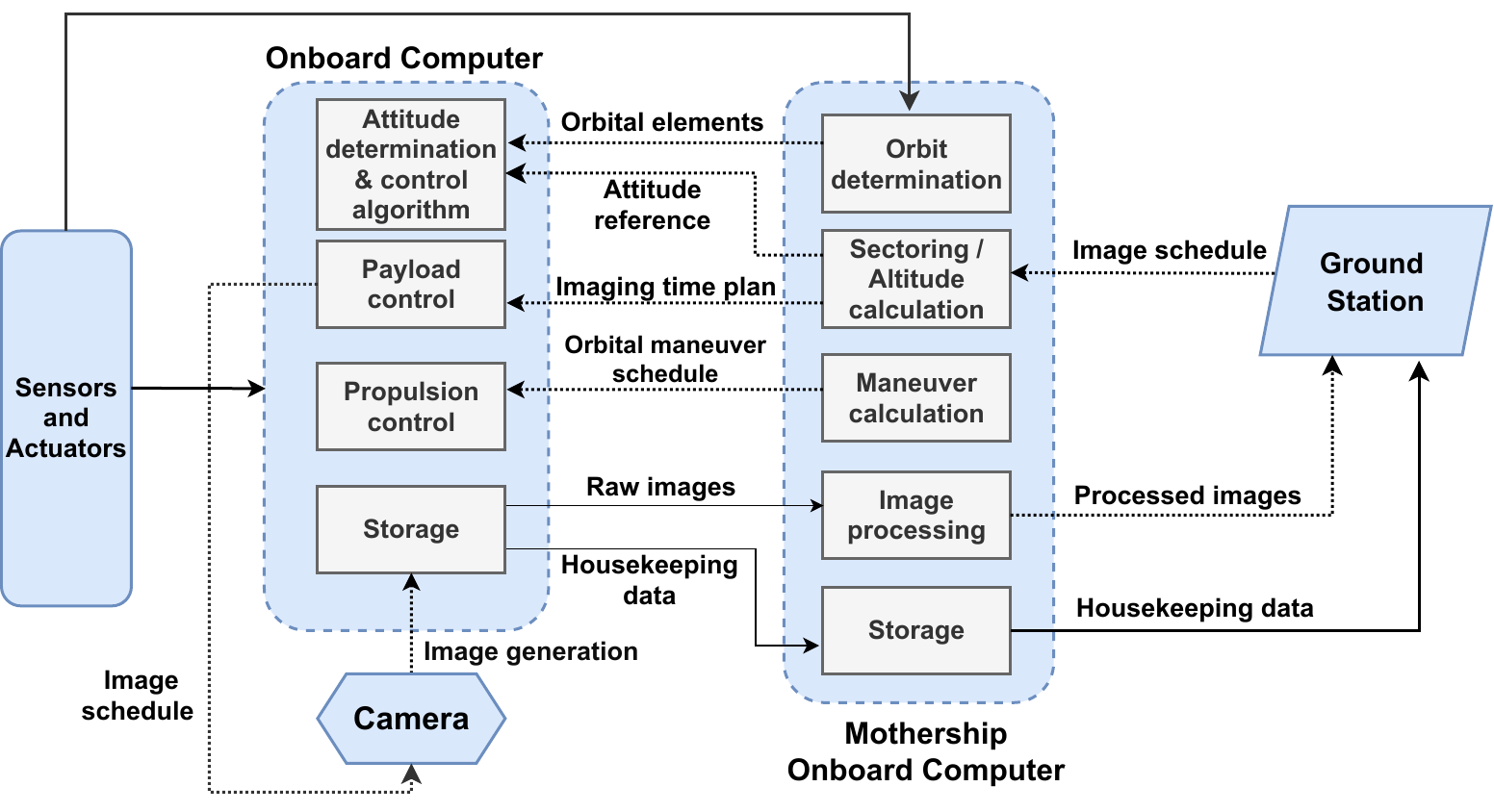}
\caption{ }
\label{figure_leo_nat}    
\end{subfigure}
\caption{Decomposition of the systems for each use case: (a) Decomposition of the EO through a single LEO satellite; (b) Decomposition of the EO through a LEO satellite network.}
\end{figure*}

\subsection{EO through a Single LEO Satellite}


The workflow for a single LEO satellite starts with the ground station sending a manually or automatically scheduled command containing image schedule data. This data flow is received by the payload control system, which directs the camera to capture images through the image generation process. The captured images are then stored temporarily in the storage component. Raw images from the storage are sent to the image processing module. Processed images and housekeeping data are returned to the ground station for analysis. The attitude determination and control algorithm maintains correct orientation and stability. In contrast, the payload control system coordinates image support maneuvers through the propulsion control component to adjust the satellite’s position and orientation as necessary, ensuring optimal image capture. Sensors and actuators provide feedback to both payload control and propulsion control systems. All these components described above are depicted in Fig. \ref{figure_single_leo} and form a cohesive system, facilitating the LEO satellite's ability to collect, process, and transmit data directly to the ground station.


\vspace{-0.2cm}

\subsection{EO through a LEO Satellite Network}

The workflow for a LEO satellite network for EO begins with the ground station sending an image schedule data flow, which can be scheduled manually or automatically. The mothership's onboard computer receives the flow, coordinating various functions, including orbit determination, sectoring/altitude calculation, and maneuver calculation. The imaging time plan and orbital maneuver schedule are sent to the LEO satellite's onboard computer. On the satellite, the payload control system directs the camera to capture images through the image generation data flow. Captured images are temporarily stored in the storage component. Raw images and housekeeping data are transmitted to the mothership's onboard computer for processing. The image processing component processes the raw images. The processed images and the housekeeping data are then transmitted to the ground station for analysis, and a copy is stored in the mothership. The detailed depiction of these components and their interactions can be seen in Fig. \ref{figure_leo_nat}.

\section{Attack Surfaces and Techniques} \label{attacks}

\begin{table}[!tbh]
    \caption{Attack surfaces outline for a single LEO satellite.}
    \renewcommand{\arraystretch}{1.5} 
        \centering
         \scalebox{0.75}{
    \begin{tabular}{|p{2cm}|p{1.4cm}|p{2cm}|p{2cm}|p{2cm}|}
        \hline
        Component & Input & Processing & Output & Related data flow \\ \hline\hline
        Ground station & Processed images & Image service & Image product & Processed images  \\ \hline
        Image processing & Raw images & Image service & Processed images & Processed images  \\ \hline
        Payload control & Image schedule & Scheduling service & Scheduled commands & Image schedule \\ \hline
       Camera & Image schedule command & Image acquisition service & Image data & Image generation  \\ \hline
        Storage & Image data & Data storage service & Raw images & Image generation  \\ \hline
        Image processing & Raw image data & Image processing service & Processed images & Processed images  \\ \hline
    \end{tabular}}
    \label{tab:attack_surface_single_leo}
\end{table}

The attack surface for our link capabilities was defined and aligned with the engineering scope. The component inventory, related input, related processing, and related output with data flow were used to identify relevant adversarial techniques. This approach allowed the demonstration of how to apply SPARTA as a threat source during the application of the technical security engineering process. Space-focused adversarial capabilities continue to evolve, especially in military operations. EO missions must be engineered with intrinsic security and a military-hardened mindset. This is especially important based on the convergence of commercial operators into a hybrid military landscape for space operations. The commercial space data system attack surface area has become a military surface area, so commercial security \textit{becomes} military security. In addition, adversaries are constantly seeking exploitation attack surfaces further left, software-focused, in both commercial and hybrid government space data systems. Accelerating the ``shift left" mindset for commercial, military, and other critical space data system operations requires a paradigm shift. This shift, component-level \textit{secure-by-design}, represents the attack surface reduction military advantage needed to protect national and allied assets.  


\subsection{EO through a Single LEO Satellite}







In our threat enumeration, we employ a strategic approach that involves mapping the sequence of operations to the most critical attack techniques delineated by the SPARTA framework. However, we first identify our attack surfaces by decomposing the inputs, processing, outputs, and data flow for each low-level component, as presented in Table~\ref{tab:attack_surface_single_leo}. 



\paragraph{Ground-to-Space Trust - [SPARTA Attack Technique: IA-0009 Trusted Relationship]} 




The trust relationship established between a ground station component and a payload control element on the satellite in the uplink consists of the image schedule data flow, which is crucial for maintaining integrity and resilience. An adversary could exploit this trust, gain unauthorized access, and inject incorrect commands and schedules. This could lead to manipulating the timing or content of scheduled image captures. Alternatively, they could prevent uplink data transmission by disrupting the command flow, preventing essential commands from reaching the payload control element, and leading to operational disruptions.



\paragraph{Image Acquisition Process - [SPARTA Attack Technique: IA-0009 Trusted Relationship]}  

The trust relationship resides within the satellite's components, including the attitude determination and control algorithm, propulsion control, sensors, and actuators. It is crucial to ensure that malicious commands or error values do not impact the lone space vehicle's ability to support the primary mission's image schedule data flow. From an adversary's viewpoint, manipulating the attitude determination and control algorithm could lead to misalignment, affecting image quality. Similarly, compromising propulsion control systems could disrupt satellite maneuverability, impacting image capture locations and schedules. Sensors and actuators, if compromised, can result in inaccurate data collection, affecting the overall mission objectives. Furthermore, the interconnectedness of these components increases the risk, as a single compromise could lead to broader system vulnerabilities and potentially grant unauthorized access to critical functions.

\paragraph{Image Storage - [SPARTA Attack Technique: IA-0006 Compromise Hosted Payload]} When considering the hosted payload, our focus extends to the components of the camera, camera storage, image processing, and housekeeping data flow. From an adversary's perspective, the camera's attributes, such as focus and scale, are vulnerable, as a compromise in technical integrity could be exploited to manipulate or distort images for deception. Additionally, once images are acquired and stored, adversaries may attempt to exploit vulnerabilities in the transmission of the housekeeping data by manipulation to simulate operational failures, aiming to deceive ground operations and compromise mission objectives.

\paragraph{Image Delivery to Ground Station- [SPARTA Attack Technique: IA-0009 Trusted Relationship and DE-0002 Prevent Downlink]} 

The trust relationship is critical in the downlink, ensuring the ground station receives EO-processed image data exclusively from the satellite's storage element. 

An adversary could prevent downlink data transmission, resulting in significant data loss and disrupting real-time monitoring and decision-making processes. This disruption would affect operational tasks reliant on timely data updates, necessitating manual intervention or rescheduling.




\subsection{EO through a LEO Satellite Network}

\begin{table}[!t]
    \caption{Attack surfaces outline for the mothership's OBC.}
    \renewcommand{\arraystretch}{1.5} 
    \centering
    \scalebox{0.9}{
    \begin{tabular}{|p{2.2cm}|p{0.7cm}|p{1.25cm}|p{0.7cm}|p{2.5cm}|}
        \hline
        Component & Input & Processing & Output & Related data flow \\ \hline\hline
        Orbit determination & \multirow{5}{*}{\begin{turn}{90}\begin{minipage}{2.5cm}Signal from payload\\ control\end{minipage}\end{turn}} & Commands & \multirow{5}{*}{\begin{turn}{90}\begin{minipage}{2.5cm}Acknowledgement\end{minipage}\end{turn}}& Sectoring and altitude  \\ \hhline{-~-~-}
        Sectoring altitude calculation &  & Commands &  & Processed images  \\ \hhline{-~-~-}
        Maneuver calculation &  & Commands &  & Processed images  \\ \hhline{-~-~-}
        Image processing &  & Image data &  & Processed images  \\ \hhline{-~-~-}
        Storage &  & Image data &  & Processed images  \\ \hline
    \end{tabular}}
    \label{tab:attack_surface_network_leo}
\end{table}



The same methodology is applied to the second scenario, with the specifics of the attack surface depicted in Table \ref{tab:attack_surface_network_leo}. As the characteristics specific to a LEO satellite within the network are similar to those in the first use case, which involves a single LEO, the sequences detailed below do not include the previous use case.


\paragraph{Primary and Secondary Trust Relationship - [SPARTA Attack Technique: IA-0009 Trusted Relationship]}  The trust relationship established with the mothership's OBC and a trailing satellite's OBC introduces a technical and operational attack surface area. The trailing satellite must rely on orbital elements, attitude reference, imaging time plan, orbital maneuver schedule, and physical proximity to maintain inter-satellite communications. 

From an adversary's viewpoint, exploiting vulnerabilities in this trust relationship could result in unauthorized access to critical satellite control systems. Manipulating orbital elements might lead to inaccuracies in positioning or trajectory deviations, affecting both the mothership and trailing satellite. Interfering with imaging schedules could compromise data collection or analysis, impacting either satellite's ability to carry out its missions effectively.




\paragraph{Space-to-Space Transmission - [SPARTA Attack Technique: IA-0009 Trusted Relationship]} The same trust relationship extends to the ISL links, which are crucial for the transmission of the orbital elements, attitude references, orbital maneuver schedules, and imaging time plans from the mothership's OBC to the satellites' OBC.

From an adversarial standpoint, manipulating orbital elements or maneuver schedules, interfering with imaging schedules, and disrupting communication channels between the mothership and satellites are potential attack scenarios. These actions could disrupt the satellite's orbit or maneuvers, risking collisions or trajectory deviations. Such interference compromises operational capabilities and mission objectives. Another concern is the disruption of imaging schedules, allowing attackers to alter image capture timing or sequence. This leads to gaps or inconsistencies in data collection and undermines critical observations or data analysis.


\section{Security Foundation} \label{Security}

This section maps the attack techniques identified above to \textit{secure-by-design} principles. The principles derived for both scenarios are summarized below:
\begin{itemize}
   \item {\emph{COMSEC (CM0002):}} ensures confidentiality and integrity through secure communication protocols, protecting sensitive data from interception and tampering;
   \item {\emph{Onboard Intrusion Detection \& Prevention (CM0032):}} implements real-time monitoring and response mechanisms on the EO OBC to detect and mitigate threats;
   \item {\emph{Segmentation (CM0038):}} divides the blocks into isolated compartments to limit the spread of intrusions and protect critical payload and control components;
   \item {\emph{Least Privilege (CM0039):}} restricts access rights for processes to the minimum required for their functions;
   \item {\emph{Robust Fault Management  (CM0042):}} incorporates comprehensive fault detection, diagnosis, and recovery mechanisms to maintain system functionality;
   \item {\emph{Alternate Communications Paths (CM0070):}} establishes redundant communication paths to ensure continuous data transmission and command control in case of primary path failures.
\end{itemize}




Following this outline, we can draft the \textsc{shall} statements that define the essential countermeasures required. The example below illustrates a \textsc{shall} statement specific to securing ISL links that, although directly applicable to our second scenario, addresses broader ISL link security challenges we've identified. Engineering teams must receive all \textsc{shall} statements for every data flow to ensure a comprehensive and secure system design. These \textsc{shall} statements provide engineering teams with actionable guidance framed in engineering language that incorporates threat mitigation and resilience considerations, serving as foundational input for the overall systems engineering process.


\begin{enumerate}[label={\bfseries EO \arabic*:},leftmargin=*]
 \item {Attitude determination and control algorithm block} \textit{shall} {implement onboard intrusion detection mechanisms}:
      \begin{enumerate}[label={\bfseries \quad - EO \arabic{enumi}.\arabic*:},leftmargin=2.5em]
          \item to monitor unauthorized and/or malicious access attempts;
      \end{enumerate}
 \item {Attitude determination and control algorithm block} \textit{shall} {implement segregation}:
      \begin{enumerate}[label={\bfseries \quad - EO \arabic{enumi}.\arabic*:},leftmargin=2.5em]
          \item to ensure that control algorithms are isolated from other system components to prevent cross-contamination of faults;
          \item to provide multi-layered security for critical algorithms.
      \end{enumerate}
 \item {Payload control block} \textit{shall} {establish and maintain alternate communication paths}:
      \begin{enumerate}[label={\bfseries \quad - EO \arabic{enumi}.\arabic*:},leftmargin=2.5em]
          \item to ensure data transmission continuity in the case of primary link failure;
      \end{enumerate}
 \item {Propulsion control block} \textit{shall} {feature robust fault management systems}:
      \begin{enumerate}[label={\bfseries \quad - EO \arabic{enumi}.\arabic*:},leftmargin=2.5em]
        \item to detect, analyze, and promptly rectify propulsion system anomalies;
        \item to support fallback operational modes that can be activated during fault conditions to maintain basic functionality.
      \end{enumerate}

\end{enumerate}




    



\section{Conclusion} \label{sec:conclusion}

In this paper, we presented two scenarios for EO missions, revealing that these missions face numerous threats, with signal integrity and resilience emerging as critical vulnerabilities. The \textit{secure-by-component} approach was instrumental in reaching such findings; by applying this procedure, we have developed secure blocks with precise specifications. Our priority is to prevent the permanent or temporary loss of satellite control during attacks, achieved through robust intrusion detection, prevention mechanisms, and fault management protocols.

  
 

\bibliographystyle{IEEEtran}
\bibliography{citations}

\end{document}